\documentclass{llncs}
\usepackage{vdm-rg}
\usepackage[final,bookmarks,bookmarksopen,bookmarksdepth=2]{hyperref}
\usepackage{pdflscape}
\usepackage{graphicx}
\leftRecord
\leftCases

\def\ProofIndent{0em}
\renewcommand{\VDMInnerIndent}{\quad}
\newcommand{\I}[1]{\hbox to #1em{}}
\DeclareSymbolFont{largesymbols}{OMX}{yhex}{m}{n}
\DeclareMathAccent{\wideparen}{\mathord}{largesymbols}{"F3}
\newcommand{\posvals}[1]{\wideparen{#1}}

\newMonadicOperator{\be}{be}
\newMonadicOperator{\st}{st}
\begin{document}

\title{Rely/Guarantee, Refinement and the ABA Problem: Part 1}
\author{Nisansala P. Yatapanage}
\institute{School of Computing, The Australian National University}

\maketitle

%

\begin{abstract}
Rely/guarantee reasoning provides a compositional way of reasoning about concurrency. The ABA problem occurs in many non-blocking concurrent data structures, where a change made by a concurrent process may be undetected by other processes. Guarantee conditions provide a useful mechanism for reasoning about such changes, as is demonstrated by two non-blocking examples, the Treiber stack and the Herlihy-Wing queue. The ABA problem can be identified by the program making a step where the \textit{before} and \textit{after} states do not correspond to a valid step at the sequential level. Therefore, such invalid behaviour relates to a failure of the guarantee condition. As such behaviour is non-\textit{linearisable}, this suggests a strong relationship between refinement with rely/guarantee and linearisability.
\end{abstract}


\section{Introduction}\label{Intro}
Concurrent algorithms are difficult to verify due to the interference between the processes. Non-blocking algorithms are notoriously difficult, as the processes are not required to wait for the other to complete a task. The rely/guarantee approach \cite{Jones83a,Jones83b} permits compositional reasoning about concurrent processes, by providing a means for describing the interference between the processes. A rely condition is a relation that describes the interference that the process can tolerate from the environment. Its counterpart, the guarantee condition, is a relation describing what the process will ensure to do. The guarantee condition must hold between the initial and final states of a program step or a sequence of program steps, as it is a transitive relation. Similarly, the rely condition must hold between the initial and final states of an environment step or sequence of environment steps. In order for two processes to execute concurrently, the guarantee of each must satisfy the rely of the other.

The rely/guarantee technique has been used to verify non-blocking algorithms such as Simpson's Four Slot \cite{PVEaCfC}. There are, however, a class of algorithms for which it is particularly challenging to find suitable rely and guarantee conditions that capture the intended behaviour of the system. For example, in \cite{JonesYatapanage19}, an attempt was made to verify a concurrent garbage collecting algorithm, but it was found to require additional constructs such as ghost variables in order to achieve the desired result. This paper examines two small but challenging algorithms, for which it is difficult to use standard rely/guarantee conditions, but where the use of rely/guarantee still offers useful benefits and interesting insights into the behaviour of these systems. 

In particular, these two examples are often used to demonstrate verification approaches that prove \textit{linearisability}. Linearisability \cite{linearisability} ensures that a concurrent program can only exhibit behaviour that matches a sequential abstract specification, by treating each concurrent operation as if it occurs at a single point in its execution, the \textit{linearisation point}. The trace of concurrent behaviour should match with an abstract trace, with respect to the linearisation points. This is explained further in Section \ref{lin}.

The examples discussed in this paper both demonstrate an interesting feature: incorrect (non-linearisable) behaviour can be identified by considering just the single program steps of an operation. This suggests that it is possible to reason about these systems in terms of guarantee conditions that hold over single program steps, rather than needing to examine the entire trace. However, relating the concurrent specification to the abstract sequential specification still presents challenges, as will be seen. The difficulty is that very little can be stated about what changes the environment has made to the variables. 

The goal of the paper is not to present a fully-developed approach, but instead to work through the examples, exploring what issues arise and the implications they have for verifying such systems. The examples investigated are the well-known Treiber stack, a standard algorithm used for discussing linearisability, and the Herlihy-Wing queue, which is of interest due to its non-fixed linearisation points. In the Herlihy-Wing queue, the linearisation points can only be determined by considering the future actions of environment processes.

The rest of the paper is structured as follows:  Section \ref{Treib} explores the first example, the Treiber stack, along with the insights gained from specifying rely/ guarantee properties for it, Section \ref{Queue} discusses the second example, the Herlihy-Wing Queue, Section \ref{Related} discusses related work and Section \ref{Conclusion} gives the conclusions.

\section{The Treiber Stack} \label{Treib}
The Treiber stack \cite{Treiber86} is a concurrent non-blocking stack data structure. It permits the standard stack operations \textit{push} and \textit{pop} to run concurrently, without forcing any process to wait until the other has finished. It makes use of the \textit{Compare-and-Swap (CAS)} operation, a low-level hardware primitive that is available on most systems, which allows an update to be atomically performed on the condition that the variable's current value is equal to a given value. For example, $CAS(x,y,z)$ will atomically check whether variable $x$ is equal to $y$, and if so, $x$ will be updated to $z$. The CAS returns true if it succeeds and false if it does not. 

The code in Fig \ref{Tcode} shows the basic idea of the \textit{push} and \textit{pop} operations, though this version suffers from the $ABA$ problem, described in Treiber's original paper \cite{Treiber86}, where a change is undetected because the original head node has been returned to the stack. This was resolved by Treiber using a $DCAS$, which atomically checks both the variable to be updated and also a count keeping track of further changes, but this paper assumes there is garbage collection, which also resolves the issue. 

\begin{figure}
\begin{verbatim}
push(v){
   Node n = new Node(v);
   do{
     x = head;
     n.next = x;
   }while(!CAS(head, x, n));
}

pop(){
  do{
    x = head;
    if (x == null) return null;
    y = x.next;
    v = x.val;
  }while(!CAS(head, x, y));
  return v;
}
\end{verbatim}
\caption{The code for the Treiber stack}
 \label{Tcode}
\end{figure}

The $push$ operation creates a new node with the given data value, records the current value of the list's head,  and then sets the new node's next pointer to point to the current head. It then uses a $CAS$ operation to attempt to make the new node become the new head. The $CAS$ is used to ensure that no other process has changed the list in the meantime. It checks that the head still matches the value recorded earlier, and only if so, it proceeds with the change to the head. If the $CAS$ fails, the operation starts again, recording the new state of the head. 

The $pop$ operation works in a similar manner. It notes the current value of the head. If it is null, the operation returns null. If not, the value of the head is noted and the next node after the head is also noted. The operation then uses a $CAS$ to attempt to set the head to the node after the head. Again, a $CAS$ is used to ensure that no other process has modified the list in the meantime. It succeeds only if the head has not been changed from the value noted earlier. If it has, the operation starts again.

\begin{figure}
  \includegraphics[width=\linewidth]{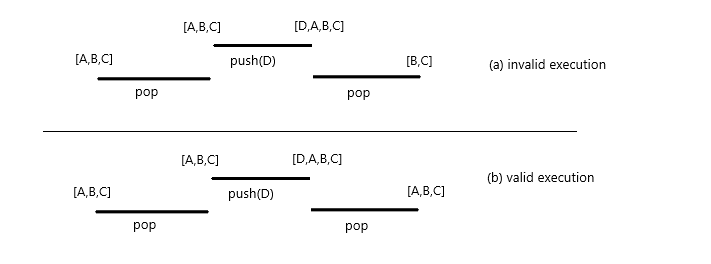}
  \caption{Valid and Invalid execution traces of the Treiber stack}
  \label{stackTraces}
\end{figure}
      
Fig. \ref{stackTraces} illustrates why the $CAS$ operations are necessary. Fig. \ref{stackTraces}(a) shows the case where one process does a $pop$, but before it finishes, another process executes $push(D)$. The first process then continues and sets the head to $B$, which it had earlier noted was the next node after the head, effectively deleting both $A$ and the new pushed node $D$. This situation is avoided using the $CAS$ statements. Fig. \ref{stackTraces}(b) shows the valid version of the same trace, where the $pop$ correctly removes only node D.

\subsection{Linearisability}\label{lin}
The Treiber stack is a standard example used for showing the concept of \textit{linearisability}. Linearisability \cite{linearisability} is a correctness condition that requires that a set of concurrent concrete processes only exhibit behaviour that matches an abstract specification, by considering each concrete operation as effectively occurring at a specific point in its operation. This point is known as the \textit{linearisation point} of the operation. For example, in the Treiber stack, the linearisation point of the $push$ operation would be the CAS statement. For the $pop$ there are two linearisation points: the CAS statement and the statement that returns null in the case of an empty stack. Linearisability considers \textit{histories} of events. The invalid behaviour shown in Fig. \ref{stackTraces}(a) is not linearisable to an abstract specification, because if the $push$ process has passed its linearisation point before the $pop$ process reaches its linearisation point, then it corresponds to an abstract history where the $push$ occurs before the $pop$. However, there is no such abstract history with a final list state of $[D,A,B,C]$, as the $push$ would have added $D$ and the $pop$ would have then removed $D$. 

Proving linearisability for a given concurrent program is usually accomplished using techniques such as simulation (see \cite{dongolLinear} for a comprehensive survey). As will be seen in the following sections, rely/guarantee conditions could provide an alternative formalism for verifying linearisability by specifying the linearisability requirements as part of the rely/guarantee conditions.     

\subsection{Rely/Guarantee Conditions for the Treiber Stack}
Specifying rely/guarantee conditions for the Treiber stack is not straight-forward. The issue is that a process has no way of knowing what changes another process may have done to the stack. The $push$ operation does not require anything from the environment processes except that they should respect the changes it made. This doesn't prevent the environment processes from removing the nodes pushed by this operation; it is valid for a concurrent $pop$ operation to remove a pushed node in the next step. However, a concurrent $pop$ should not ignore the fact that this $push$ added a node to the stack, i.e. it should not produce the result shown in Fig. \ref{stackTraces}(a). Similarly, this $push$ should not ignore any changes made by other $push$ or $pop$ operations running concurrently with it. Therefore, the rely conditions for $push$ and $pop$ need to somehow express this property, but without unnecessarily constraining the behaviour of the other processes.

At the abstract level, the state is defined as follows, using VDM notation (see \cite{Jones90a} for an introduction):
\begin{record}{\Sigma\sb{0}}
list: \seqof{Val}
\end{record}

\noindent It simply contains a sequence of values. At this level, the $pop$ and $push$ operations execute atomically. There is therefore no need of rely and guarantee conditions, as there is no interference. The effects of the operations are specified entirely in the post conditions.

The $push$ operation takes a value $v$ and inserts it as the new head of the list.

\begin{op}
\opname{push\sb{0}}
\args{v:Val}
 \pre{\true}
 \post{list' = \seq{v} \sconc list}
 \end{op}
 
The $pop$ operation has no parameters and returns a value $x$, the original head of the list.

\begin{op}
\opname{pop\sb{0}}
\args{}
\res{x:Val}
 \pre{\true}
 \post{list' = \tl list  \And x' = \hd list}
 \end{op}

At the next level of abstraction, the operations are no longer assumed to be atomic, but still operate on an abstract sequence. Rely and guarantee conditions are necessary to describe the interference at this level.

\begin{record}{\Sigma\sb{1}}
list: \seqof{Val}
\end{record}

\noindent A first attempt at the rely for the $push$ might be to state that any environment steps either left the list's head unchanged or set it to the second node in the list: \\

\textbf{$rely-push\sb{1}$}: $(list' = list) \Or (list' = \tl list)$\\

However, this would constrain the environment behaviour to be a single $pop$, whereas it is valid for several $pop$ operations to be performed concurrently with this $push$, as well as for other $push$ operations to execute. There is, in fact, nothing at all that can be said about the state of the list after environment steps have completed. The list may even be left completely empty, or containing a whole list of new nodes pushed by other processes.  

Before continuing, first note that the post conditions of both the $pop$ and $push$ operations are also impossible to state, because the final state of the stack is unknown. The \textit{possible values} notation \cite{PVEaCfC} has been used effectively to state properties of this nature, but even it is insufficient in this case. The possible values notation $\posvals{x}$ means the set of values that $x$ held during the execution of the operation. Using possible values, it can be stated that the list contained a given node \textit{at some state} during its operation. This would allow the post condition of $push$ to state that the value to be pushed was added at least at some point.\\

\textbf{$post-push\sb{1}$}: $\exists {l \in \posvals{list}}{\hd l = v}$\\

However, this post condition is too weak, as it does not prevent the $push$ operation from making further changes to the list, which it is not supposed to do. This has to be constrained using the guarantee condition.

\subsubsection{Guarantee conditions for Specifying Linearisability Requirements}

Return to Fig. \ref{stackTraces}(a), which was one of the cases required to be prevented. Notice that before the $pop$ erroneously removed both nodes D and A, there had to be a step of $push$ that added node D. If there hadn't been such a step, then the behaviour of the $pop$, which assumed the head to still be A, would have been correct. Considered from the point of view of the $pop$, there was an environment step which changed the stack to $[D,A,B,C]$, followed by its own program step (or multiple steps) which changed the stack to $[B,C]$. In other words, if the state of the stack is considered just before and after the $pop$'s program steps, its behaviour would be considered invalid, as it cannot change a stack from $[D,A,B,C]$ to $[B,C]$. This is exactly what a guarantee condition provides: the ability to reason about the state of the system before and after a sequence of program steps. 

The only task remaining is to construct a suitable guarantee condition that ensures the invalid behaviour discussed above is prevented. The abstract post condition of $pop$ is $list' = \tl list \And x' = \hd list$, as the change occurs atomically in one step. On the concrete level, if the $pop$ operation passes its linearisation point on a particular step, then it should match the behaviour of the abstract post condition. Otherwise, the list should be unchanged, i.e. $list' = list$. Therefore, the concrete guarantee must show that each step either matches the abstract post condition or leaves the list unchanged: \\

\textbf{$guar-pop\sb{1}$}: $list' = list \Or (list' = \tl list \And x' = \hd list)$ \\

Since guarantee conditions are required to be transitive, this prevents two removals within one sequence of program steps. However, it does not prevent two removals occurring with environment steps in between. 

Therefore, it is clear that the linearisation point needs to be considered. Steps that do not contain the linearisation point need to leave the shared variables unchanged, i.e. preserving $list' = list$, whereas steps containing the linearisation point should be shown to either achieve the abstract post condition or leave the list unchanged, i.e. $list' = list \Or list' = \tl list$. The option to leave the list unchanged is still required, as not all of the steps within this sequence of program steps will update the list, and the guarantee must hold transitively over all the steps.

To prevent multiple removals, a flag is needed that indicates whether or not a removal has occurred. If the flag is true, no further removals are possible. The flag effectively models that only stuttering steps should occur before and after the linearisation point.

The full $pop$ operation is:

\begin{op}
\opname{pop\sb{1}}
\args{}
\res{x:Val}
 \pre{\true}
 \rely{\true}
 \guar{(list' \neq list \implies list' = \tl list \And flag' = true \And flag = false) \And \\
 	(list' = list \implies flag' = flag)}
 \post{\true}
 \end{op}

The proof obligation now is to show that $post-pop\sb{0}$ (the post condition of the sequential version) is refined by the program and environment steps executing under $guar-pop\sb{1}$ and $rely-pop\sb{1}$ respectively, followed by $post-pop\sb{1}$. However, this is where the difficulty lies. There is nothing that can be stated about the behaviour of the environment, as the environment may include further $pop$ or $push$ operations, which could even remove all the existing elements in the list and insert new ones. Without knowing exactly what the environment has done to the rest of the list, it is impossible to conclude that $list' = \tl list$ at the end of the operation. 

On the sequential level, environment steps can only occur before or after the operation. Therefore, to relate it to the concurrent level, instead of comparing the values of $list$ at the start and end of the concurrent operation to the start and end of the atomic level, the values of $list$ before and after the linearisation point are needed. A possible solution may be to introduce a way of recording the state before and after the step where the flag is set to true. Further investigation is needed about this.

\begin{op}
\opname{push\sb{1}}
\args{v:Val}
 \pre{\true}
 \rely{\true}
 \guar{(list' \neq list \implies list' = \seq{v} \sconc list \And flag' = true \And flag = false) \And \\
 	(list' = list \implies flag' = flag)}
 \post{\true}
 \end{op}
 
The $push$ operation, given above, has a similar structure and the same issues are encountered when attempting to prove that it refines the atomic $push$. From here, the next level of refinement would be closer to the code level, with a structure such as a linked list to replace the sequence. To save space, this refinement step has been omitted as it does not present any further interesting insights.

\section{The Herlihy-Wing Queue} \label{Queue}
The previous example contained \textit{fixed} linearisation points. Some problems contain linearisation points that cannot be determined ahead of time, as they depend on the execution of the other processes. Refer to \cite{dongolLinear} for a comprehensive comparison of different approaches for problems with different types of linearisation points. The Herlihy-Wing Queue \cite{linearisability} is an example of a problem with linearisation points that cannot be determined from the start. This makes it an interesting and challenging problem for showing that linearisability holds.

The queue is an array-based structure that allows concurrent enqueue and dequeue operations. The code for each is given in Fig. \ref{HWcode}. 
\begin{figure}
\begin{verbatim}
enq(v){
  <index = last; last = last + 1>;
  q[index] = v;
}

deq(){
 while(true){
   range = last;
   index = 0;
   while(index < range){
     x = null;
     swap(q[index],x);
     if (x != null) return x;
     index++;
   }
 }
}
\end{verbatim}
\caption{The code for the Herlihy-Wing Queue}
 \label{HWcode}
\end{figure}

The statements inside the angular brackets in $enq$ execute atomically; the operation sets its own local index to $last$ and increments $last$ in one atomic step. It then stores a value into the slot with that index. The $deq$ continuously traverses the array, searching for the first non-null node it encounters. It does this by atomically swapping the value at the current index with a null value. If the value that was read from that slot, $x$, is not null, this value is returned. Otherwise, it continues searching. This process is potentially non-terminating, if nothing is added to the queue, or if other $deq$ operations running concurrently always take the items first.

The difficulty of this problem is that the linearisation point for the $enq$ depends on the order of execution of the other operation. If the $deq$ has already checked the first slot (index = 0) by the time the $enq$ inserts a value, and then a second $enq$ inserts a value into the second slot (index = 1), the $deq$ will return the second value before the first, which appears to break the FIFO ordering of the queue.

The following is an example execution trace that results in this situation, where a single $deq$ process is interleaving with two $enq$ processes:\\\T6
$deq$ (checks slot 0) \T6
$enq(A)$ (inserts into slot 0) \T6
$enq(B)$ (inserts into slot 1) \T6
$deq$ (checks slot 1 - returns B)\\

As explained in \cite{dongolLinear}, this concrete execution is actually equivalent to an abstract sequence where the second value B was enqueued before the first value A. This is what creates complications for verifying linearisability. 

In this case, at first it would appear that the previous section's technique of examining the state of the queue before and after a single step does not work. There is a $deq$ step where the queue contains A followed by B, i.e. [A, B], before the step and [A,-] afterwards, which does not respect FIFO ordering. However, unlike for the Treiber stack, this \textit{could} be considered acceptable behaviour, if the $deq$ had already checked the first slot before A had been inserted. On the other hand, traces where the $deq$ removed B first even though A had been there before the $deq$ operation started should be disallowed. In other words, the transition from [A, B] to [A,-] is only valid if the current index value of the $deq$ was already equal to $1$ at the start of this sequence of $deq$ steps. 

Therefore, the pattern from the last section still holds if the states of both the queue and the local index variable are considered. Invalid behaviour is where a state with [A,B] and $index = 0$ transitions to a state with [A,-], whereas a state with [A,B] and $index = 1$ can validly transition to [A,-]. In the next section, these ideas will be used to devise an appropriate guarantee condition.

\subsection{Rely and Guarantee Conditions for the Queue}

At the abstract level, the state is defined the same way as for the Treiber stack, as a sequence of values:
\begin{record}{\Sigma\sb{0}}
list: \seqof{Val}
\end{record}

The $enq$ and $deq$ operations are very similar to the $push$ and $pop$ of the Treiber stack, with the only difference that $enq$ inserts items at the opposite end of the list than where $deq$ removes them from.

\begin{op}
\opname{enq\sb{0}}
\args{v:Val}
 \pre{\true}
 \post{list' = list \sconc \seq{v} }
 \end{op}
 
\begin{op}
\opname{deq\sb{0}}
\args{}
\res{x:Val}
 \pre{\true}
 \post{list' = \tl list \And x' = \hd list}
 \end{op}

%
%
%

The next level is designed to model the code given in Fig. \ref{HWcode}. While it is possible to use an intermediate level of abstraction that exhibits concurrency using the same abstract sequence representation, this only presents the same issues as for the Treiber stack, so is not shown here. The more concrete level closer to the code presents some further interesting challenges. This level contains a sequence of values, an index, a variable $n$ representing the length of the array and a variable $last$ representing the index of the last item inserted into the queue. Note that the concrete level could have been described at the heap level, as was done in \cite{JonesYatapanage-15}, instead of modelling the array as a sequence. However, this would not make any difference for the points discussed in this paper, so this model has been used to simplify the presentation. 

The invariant ensures that $index$ and $last$ remain within the boundaries of the array.  \footnote{In Herlihy and Wing's original presentation \cite{linearisability}, the array was infinite. In this version, a finite array is used to prevent potential non-termination of some functions shown later in this section. To remain close to the original code, it is assumed to stop enqueuing when the end of the array has been reached (rather than treating it as a circular array). Appropriate checks in the code have not been shown but it is easy to assume what they would be.} 

\begin{record}{\Sigma\sb{1}}
q: \seqof{Val}  \\
index: \Nat \\
n: \Nat \\
last: \Nat
\end{record}
\where
\begin{fn}{inv-\Sigma\sb{1}}{q, index, n, last} \\
\signature{\seqof{Val} \x \Nat \x \Nat \x \Nat \to \Bool}
0 \leq index \leq n \And 0 \leq last \leq n
\end{fn}

\noindent The specification for $deq$ is:\\

\begin{op}
\opname{deq\sb{1}}
\args{}
\res{x:Val}
 \pre{\true}
 \rely{flag = false \And index \neq \nil \implies q'(index) = q(index)}
 \guar{(flag = true \implies flag'= true \And noChanges(\sigma, \sigma')) \And \\
 (noChanges(\sigma, \sigma') \And flag = false => flag' = false) \And \\
 (\forall {i:index \ldots (index' \minus 1)}{q(i) = null})\And \\
(q(index') \neq q'(index') \implies (q'(index') = null \And flag' = true)) \And \\
(\forall {j:0 \ldots n}{j \neq index' \implies q(j) = q'(j)}}
 \post{\true}
 \end{op}
 
 where the function $noChange$ indicates that no changes were made to either an element in the queue or the $last$ variable:
 
 \begin{fn}{noChange}{s, s'} \\
\signature{\Sigma\sb{1} \x \Sigma\sb{1}}
(s.last = s'.last) \And
(\forall {i:0 \ldots s.n)}{s.q(i) = s'.q(i))}
\end{fn}

The first two clauses about the flag and $noChanges$ ensures that changes do not occur unless the flag is false and that the flag does not change to true on stuttering steps. The third clause of the guarantee states that all the slots from the original $index$ up to just before the final $index'$ should have been empty slots at the time when this sequence of program steps began. This prevents the operation from ignoring slots containing elements. Using this guarantee, a state with [A,B] and $index = 0$ could not transition to a state with the index as $1$. The fourth clause ensures that if the slot at $index'$ has been changed, it has been made null (the item was dequeued), to specify the intended behaviour of the $deq$. Finally, the last clause states that the remainder of the queue is left unchanged. 

The next step is to relate this concrete guarantee to the abstract specification. The abstract $deq$ operation which executes atomically has the post condition $list' = \tl(list) \, \And \, x' = \hd list$. In the case where the index was 1 before the guarantee step, although the concrete queue contains [A,B], it does not actually refine an abstract queue [A,B]; in fact, the corresponding abstract queue is [B,A], where the A element was inserted after the B. This is the essence of this problem: the abstract state that corresponds to a given concrete state depends on both the index value and the queue state. The concrete queue [A,B] with $index = 0$ corresponds to an abstract queue of [A,B], whereas a concrete queue [A,B] with $index = 1$ corresponds to an abstract queue of [B,A]. This is seen when attempting to find an abstract trace that is linearisable to the concrete one. \cite{dongolLinear} gives an abstract trace where the second value is enqueued first, to correspond with a concrete trace that returns the second value first. The abstract trace would be: 
$\{enq(B), enq(A), deq()\}$. 

This suggests the following retrieve function, for finding the corresponding abstract state for a given concrete state:\\
\begin{fn}{retr-deq}{(q, index, n, last)}
 \signature{\Sigma\sb{1} \to \Sigma\sb{0}}
	q(index) \sconc [q(first(s,0,n)) \ldots q(index \minus 1)] \sconc [q(index+1) \ldots q(n))]
\end{fn}

where the function $first$ returns the index of the first element that is non-null in the sequence:

\begin{fn}{first}{q, x, n}
\signature{\seqof{Val} \x \Nat \x \Nat \to \Nat}
  \If x = n
   \Then n
    \Else 
    {\If q(x) = null
    \Then first(q, x+1)
    \Else x
    \Fi}
    \Fi
\end{fn}

If there are any elements before the element at the $index$ slot, these were effectively inserted afterwards. This is reflected in the abstract queue. The retrieve function $retr-deq$ states that the corresponding abstract queue is composed of the element at the $index$ slot, followed by any non-null elements from before it, followed by the remainder of the array. For example, if the concrete queue is [A,B,C] and $index$ is 1, then a concrete $deq$ step could result in a queue of [A,-,C]. The [A,B,C] queue prior to the step corresponds to an abstract queue of [B,A,C] using $retr-deq$, while the concrete queue of [A,-,C] after the step corresponds to an abstract queue of [-,A,C]. A transition from [B,A,C] to [-,A,C] satisfies the abstract post condition of $list' = \tl list$, but only when considering the state of the queue just before and after the linearisation step. The same issue arises here as for the Treiber stack; it is not possible to conclude that the abstract atomic post condition is maintained over the whole operation of the $deq$, because other concurrent $deq$ and $enq$ operations could have changed the queue. Again, a mechanism is needed here for noting the value of the queue before and after the step where the flag became true.  

The specification of $enq$ is:\\
\begin{op}
\opname{enq\sb{1}}
\args{v:Val}
 \pre{\true}
 \rely{flag = false \And index \neq \nil \implies q'(index) = q(index)}
 \guar{(flag = true \implies flag'= true \And noChanges(\sigma, \sigma')) \And \\
 (noChanges(\sigma, \sigma') \And flag = false => flag' = false) \And \\
 (last' \neq last \implies last' = last + 1 \And index' = last \And \T4 setInd' = true \And setInd = false) \And \\
 (setInd \implies (\forall {i:0 \ldots n}{i \neq index \implies q'(i) = q(i)}) \And \T3
 	((q'(index) = v \And flag' = true \And index' = index) \Or \T3 (noChanges(\sigma, \sigma') \And index' = index))) \And \\
(setInd = false \implies ((\forall {i:0 \ldots n}{i \neq last \implies q'(i) = q(i)) \And \T3
	((q'(last) = v \And flag' = true \And setInd' = true \And \T6 index' = last \And last' = last + 1) \Or \T3
		(noChanges(\sigma, \sigma') \And index' = index)))
 }}
 \post{\true}
 \end{op}
 

Note that the $index$ variable here is local to the $enq$ and not the same as the $index$ that is local to $deq$. The first two clauses are the same as for $deq$. The third clause ensures that if $last$ is updated, which occurs when the $enq$ operation has chosen a slot to use, then that slot was $last$ and also $last$ has been incremented. The variable $setInd$ is used to indicate that $index$ has been set, to ensure that it will not be changed again. The fourth clause is for handling the case where $index$ was already updated on a previous step, leaving the option to either make no changes on this step or add the new item to the queue. The final clause handles the case where the $index$ has not been set on a previous step. On this step, both $index$ is set and the new item is added, or no changes occur. (The third clause already covers the case where $index$ is set without adding an item). 

Again, the same problem occurs when attempting to show that this specification refines the previous level. For example, suppose that an $enq$ operation selects slot $3$ to insert an item. Before it has inserted the item, another $enq$ selects slot $4$ and inserts an item. The second $enq$ completes its operation, but it is not possible to state that the abstract post condition of $list' = list \sconc \seq{v}$ is achieved, because by the time it finishes, the first $enq$ might have inserted its item, making the final queue contain more new items than just the second $enq$'s item.

\section{Related Work} \label{Related}

Many approaches exist for verifying linearisability. A comprehensive survey is given in \cite{dongolLinear}. Linearisability is usually associated with the relationship between concrete and abstract states. This aspect of the development of this paper is not new. However, most approaches use techniques such as simulation to relate the concrete and abstract traces. For example, \cite{Derrick08} present a mechanised proof of a concurrent stack using forward simulation conditions. Colvin and Groves \cite{iceccs05} verify an array-based queue using backward simulations between I/O Automata.

Dongol and Derrick \cite{DongolD13} use an interval-based logic to relate concrete operations to course-grained abstractions, to avoid the need for identifying linearisation points in the code. It would be interesting to compare whether the approach in this paper achieves the same underlying basis. 

Vafeiadis presents an approach using RGSep for verifying linearisability in \cite{Vafeiadis07}. RGSep combines rely/guarantee and separation logic. Similarly to this paper, his approach uses the relationship between the concrete and abstract states. However, the abstract operations are embedded into the program code. The approach requires the identification of linearisation points. Auxiliary variables are used to record the location of these points. While rely and guarantee conditions are used, the approach does not consider the use of the guarantee conditions to express the linearisable behaviour.

In \cite{TraceRefinementLinearizability17}, the relationship between trace refinement and linearisability is investigated. They show that trace refinement implies linearisability but not the other way around. In a similar way, it would be interesting to investigate whether there is a general relationship that always allows linearisability to be expressed using rely/guarantee conditions or whether it depends on the type of problem, and perhaps the type of linearisation points. 

Hayes \cite{REFINE2018} investigates the Treiber stack in the context of rely/guarantee specifications, but the focus there is on specifying progress and termination conditions.

\section{Conclusion} \label{Conclusion}

This paper has investigated the use of rely/guarantee reasoning to verify non-blocking problems, particularly ones which exhibit complex interactions where the correctness is normally ensured by proving linearisability. An interesting aspect that was identified is that the linearisability property can sometimes be re-stated in terms of a guarantee condition that must hold over single program steps (or a single sequence of program steps). The traces which are not linearisable always contain a single sequence of program steps that violate the guarantee condition. Examining the two examples shown in this paper has identified this common property, even though the two examples have different kinds of linearisation points. Future work will be to investigate whether this property holds generally, i.e. whether it is always possible to specify the linearisable behaviour as a guarantee condition of the operation. 

A key benefit of rely/guarantee reasoning is its compositional nature. It would therefore aid in verification to use rely and guarantee conditions to verify linearisability without having to consider the entire global system. Instead, showing linearisability reduces to proving that each operation satisfies the appropriate requirements according to the rely/guarantee conditions and the refinement from the abstract state. However, in order to achieve this goal, there are some remaining challenges to be addressed. It is interesting to note that both examples encounter the same issues when showing that the concurrent specification refines the abstract one, because it is not possible to state what the environment may do to the structure. It appears that it may be possible to resolve the problem by considering the state before and after the linearisation steps only, but further investigation is still required. Nevertheless, the two examples presented here have revealed some interesting aspects about the links between rely-guarantee and linearisability.

\subsubsection{Acknowledgements}

The idea of using a flag to record when a change has occurred was devised during a helpful and interesting discussion with Cliff Jones. The author would also like to thank Rob Colvin, Brijesh Dongol, Kirsten Winter and Graeme Smith for their useful comments on the ideas of the paper. The comments from anonymous reviewers on an earlier version also improved the paper.

\bibliography{parallel}

\end{document}